\documentclass[12pt]{article}
\usepackage[utf8]{inputenc}
\usepackage{amsfonts}
\usepackage{enumitem}
\usepackage{graphicx}
\usepackage{cite}
\usepackage{color}
\usepackage{amsmath,amssymb,amsthm}
\usepackage{mathrsfs}
\usepackage{slashed}
\usepackage[english]{babel}
\usepackage[margin=3cm]{geometry}

\def \be {\begin{equation}}
\def \ee {\end{equation}}
\def \bea {\begin{eqnarray}}
\def \eea {\end{eqnarray}}
\def \nn {\nonumber}

\def \rr {\raise.35ex\hbox{\small $\prime$}\kern-.17em{\mbox{\large $\imath$}}}

\def \dels {\partial\kern-.6em /\kern.1em}
\def \As {{A\kern-.5em / \kern.5em}}
\def \Ds {D\kern-.7em / \kern.5em}

\def \ks {k\kern-.5em /}
\def \ls {l\kern-.5em /}

\newcommand{\ci}[1]{}

\newcommand{\ba}{\begin{eqnarray}}
\newcommand{\ea}{\end{eqnarray}}
\newcommand{\bal}{\begin{align}}
\newcommand{\eal}{\end{align}}
\newcommand{\bay}[1]{\left(\begin{array}{#1}}
\newcommand{\eay}{\end{array}\right)}

\reversemarginpar
\newcommand{\hide}[1]{}

\makeatletter

\newlist{axioms}{enumerate}{2}
\setlist[axioms,1]{label=\textbf{A\arabic{axiomsi}.}, ref=A\arabic{axiomsi}}
\setlist[axioms,2]{label=\textbf{A\arabic{axiomsi}\rlap{\myEnumCounter{axiomsii}}.},%
                   ref=A\arabic{axiomsi}\myEnumCounter{axiomsii},%
                   align=parleft,%
                   leftmargin=0em,%
                   itemsep=1.4ex,%
                   before={\stepcounter{axiomsi}}}

\usepackage{tikz,pgf}
\usetikzlibrary{shapes}
\usetikzlibrary{calc}
\usetikzlibrary{decorations.pathmorphing}
\usetikzlibrary{decorations.pathreplacing,shapes.misc}
\usetikzlibrary{positioning}
\usetikzlibrary{arrows}
\usetikzlibrary{decorations.markings}

\tikzset{snake it/.style={decorate,decoration={snake,segment length=1.5mm, amplitude=.3mm}}}
\tikzset{biggerarrow/.style={
    decoration={markings,mark=at position 1 with {\arrow[scale=1.5]{>}}},
    postaction={decorate},
    shorten >=0.4pt}}
\tikzset{arrow at middle/.style={decoration={
    markings,
    mark=at position 0.5 with {\arrow{>}}}}}

\begin{document}

\begin{titlepage}

\begin{center}

\hfill
\vskip .2in

\textbf{\LARGE
Geometric Low-Energy Effective Action in a Doubled Spacetime
\vskip.5cm
}

\vskip .3in
{\large
Chen-Te Ma$^a$ \footnote{e-mail address: yefgst@gmail.com} and Franco Pezzella$^b$ \footnote{e-mail address: franco.pezzella@na.infn.it}
\\
\vskip 2mm
}
{\sl
${}^a$
Department of Physics and Center for Theoretical Sciences, \\
National Taiwan University,\\ 
Taipei 10617, Taiwan, R.O.C.\\
$^b$ 
Istituto Nazionale di Fisica Nucleare - Sezione di Napoli,\\
Complesso Universitario di Monte S. Angelo ed. 6,\\
 via Cintia,  80126 Napoli, Italy.
}\\
\vskip 1mm
\vspace{30pt}
\end{center}
\begin{abstract}
\vskip .2cm
The ten-dimensional supergravity theory is a geometric low-energy effective theory and the equations of motion for its fields can be obtained from string theory by computing $\beta$ functions. With $d$ compact dimensions, we can add to it an $O(d, d;\mathbb{Z})$ geometric structure and construct the supergravity theory inspired by double field theory through the use of a suitable commutative star product. The latter implements the weak constraint of the double field theory on its fields and gauge parameters in order to have a closed gauge symmetry algebra. The consistency of the action here proposed is based on the orthogonality of the momenta associated with fields in their triple star products in the cubic terms defined for $d\ge1$. This orthogonality holds also for an arbitrary number of star products of fields for $d=1$. Finally, we extend our analysis to the double sigma model, non-commutative geometry and open string theory.
\end{abstract}

\end{titlepage}

\section{Introduction}
\label{1}
Quantum gravity is expected to unify all theories of the fundamental interactions by combining quantum mechanics and general relativity. String theory is a candidate to provide a self-consistent framework for quantum gravity.  A crucial role at this aim is played by T-duality and S-duality.

String theory, whose action is defined in the two-dimensional world-sheet space, explores the target-space theory and its low-energy limit via one-loop $\beta$-functions \cite{Abouelsaood:1986gd} and $\alpha^{\prime}$ corrections.  One-loop $\beta$-functions provide the equations of motion satisfied, in the target space, by the background fields with which the string interacts and, in particular, the one associated with the graviton field, present in the spectrum of the closed string, generates the Einstein gravity equation. 

The main goal of double field theory (DFT) \cite{Hull:2009mi, Siegel:1993xq} is to manifestly incorporate  T-duality, i.e. the $O(d,d;Z)$ invariance in the target space with $d$ compact dimensions, as a global symmetry of the low-energy field theory deriving from closed strings living in a $D$-dimensional spacetime which is the product of a Minkowski $n$-dimensional flat space ${\cal M}=R^{n-1,1}$ with a $d$-dimensional torus $T^{d}$ ($n+d=D$). Then, the fields of DFT live in the product of ${\cal M}$ with a $2d$-dimensional doubled torus containing both the torus $T^{d}$, parametrized by the original compact coordinates $x^{m}$, and its dual $\tilde{T}^{d}$, parametrized by the dual coordinates $\tilde{x}_{m}$. The field content of DFT involves the metric field $g_{ij}$, the Kalb-Ramond field $B_{ij}$ $(i,j=1, \dots D)$ and a dilaton, i.e. the massless bosonic sector of the closed string.  Since these fields depend on $x^{m}$ and $\tilde{x}_{m}$ simultaneously,  DFT is expected to have gauge invariance both under diffeomorphisms on the former and dual diffeormophisms on the latter, i.e. a gauge invariance under doubled diffeomorphisms.  

DFT is still to be fully constructed. In \cite{Hull:2009mi}, an action for such theory is given only to cubic order in the fluctuations of fields around a fixed background. In this framework, the invariance under linearized doubled diffeomorphisms is based on the so-called {\em weak constraint} arising from the level matching condition of closed string field theory, i.e. $\Delta f \equiv 2\partial_{m} \tilde{\partial}^{m} f=0$, that has to be satisfied by fields and gauge parameters. The gauge parameters are the vector fields $\epsilon_{i} (x^{\mu}, x^{m}, \tilde{x}_{m})$ and $\bar{\epsilon}_{i} (x^{\mu}, x^{m}, \tilde{x}_{m})$ generating, respectively, the linearized gauge transformations on the metric tensor and the Kalb-Ramond fields. Therefore, the weak constraint requires fields and gauge parameters to live in the kernel of the second-order differential operator $\Delta$.

Subsequently, in \cite{Hohm:2010jy} a manifestly background independent  action has been constructed for the field ${\cal E}_{ij}=g_{ij}+B_{ij}$   and for the dilaton $d$. Further aspects of this action have been studied in \cite{Hohm:2015ugy}. Such formulation is the same as the generalized metric formulation under the weak constraint \cite{Hohm:2010pp} and results to be $O(D,D)$ invariant. In the case of $d$ compact dimensions this symmetry breaks to $O(d,d; {\mathbb Z})$ preserving the periodic boundary conditions, while the invariance under doubled diffeormorphisms is, in this case, based on the so-called {\em strong constraint}, i.e. a generalization of the weak constraint to any product of fields and gauge parameters. 

 So far, a non-trivial theory that is invariant under doubled diffeomorphisms without using the above mentioned constraints has not been found yet. One could try to formulate a theory in terms of fields automatically projected in the kernel of $\Delta$ through a suitably defined projector operator, the {\em star product}, which takes an arbitrary field or a gauge parameter to that kernel. This operator has to be used also in the gauge transformations in order to ensure that the gauge variations are allowed variations of the fields and shows, in general, a non-associative property that does not give any problems to the invariance of the action that, instead, could be spoiled by the non-closure of the gauge algebra. In other words, it may happen that, due to the non-closure of the gauge algebra, one could loose physical degrees of freedom from a total derivative term generated by applying a gauge transformation. The total derivative term or boundary term is also related to the global geometry of DFT \cite{Andriot:2011uh}. Furthermore, a closed gauge algebra in DFT is also important to ensure the closure of the supersymmetry algebra \cite{Hohm:2011nu, Hohm:2010xe}.

It has been already observed that DFT is expected to be a candidate for a supergravity theory with manifest T-duality and that the target space could be understood from a world-sheet theory by computing one-loop $\beta$-functions. An interesting work, in this direction,  has been to compute the one-loop $\beta$-functions in the double sigma model without using any constraints \cite{Copland:2011wx, Park:2016sbw}. 

The bosonic sigma model gives the dynamics of bosonic closed strings and of bosonic open strings satisfying Neumann and Dirichlet boundary conditions. In particular, a non-commutative geometry appears in the quantization of the bosonic sigma model with the endpoints of open strings attached to a brane, in the presence of a $B$-field. A Seiberg-Witten map in the target space also comes from a non-commutative geometry and it ensures the equivalence of a commutative description and a non-commutative description. A useful study of the non-commutative side consists in computing $\alpha^{\prime}$ corrections to all orders through the Moyal product.  Non-commutative geometry could also be understood through an $O(D, D)$ generalized metric \cite{Jurco:2013upa}, being the Dirac-Born-Infeld (DBI) theory \cite{Ma:2015yma, Ma:2014vqm} the low-energy effective field theory of the double sigma model for the open string.

As already claimed, the biggest issue of DFT is the loss of the closure of the gauge algebra without using the strong constraint. This latter means that it is always necessary to eliminate half of the degrees of freedom so one could not have any new physics beyond the ordinary supergravity theory. An interesting work \cite{Lee:2015qza} related to relaxing the strong constraint consists in using  a suitable product, the $\circ$-product, exhibiting properties of commutativity, associativity and distributivity and that allows to get the closure of the gauge algebra , but the issue is that the theory defined in this way could not reduce to the cubic action \cite{Hull:2009mi} obtained from closed string field theory.  

The main goal of this paper is to find a DFT-inspired string low-energy effective theory in the doubled spacetime, embodying in it the aforementioned star product that, differently from the $\circ$-product of \cite{Lee:2015qza}, does not exhibit associativity. In order to better understand the issues related to the non-associativity previously discussed, we first consider the case $d=1$ of a spacetime with only one doubled space-dimension. We find here that a closed gauge algebra can be obtained and that the action, uniquely determined by the gauge symmetry, results to be the sum of the ordinary supergravity theory and the dual supergravity \cite{Ma:2016vgq}.

The extension to arbitrary $d$ can be again understood from the triple product \cite{Ma:2016vgq} that again gives the orthogonality of the momenta, just like in $d=1$. The result implied by the triple products suggests that the cubic action can be obtained, up to a boundary term,  from a DFT defined through the star product without loosing any fluctuations of the fields. Imposing the orthogonality of momenta also gives a closed gauge algebra. This can be extended to an arbitrary number of fields and the results do not contradict the ones already known from closed string field theory. We also construct a double sigma model in terms of the star product. 

T-duality for non-geometric fluxes is defined by non-commutativity parameters. The transformations between commutative and non-commutativity parameters are physical transformations of DFT that are shown to be actually related by an $O(D, D)$ transformation. The star product is made commutative by imposing the orthogonality, which could maintain all the theoretical properties of DFT.

We first review the properties of the projector  in Sec.~\ref{2}, discuss the case for $d=1$ and do the extension to arbitrary $d$ respectively in Sec.~\ref{3} and Sec.~\ref{4}. A study of DFT and the double sigma model is exhibited in Sec.~\ref{5} and Sec.~\ref{6}. We discuss T-duality in a non-commutative space \cite{Kamani:2001yd} in Sec.~\ref{7}. The open string theory is shown in Sec.~\ref{8}. Finally, we discuss the results and conclude in Sec.~\ref{9}. 

\section{Properties of the Projector}
\label{2}
A role of the projector we are going to discuss in this section \cite{Hull:2009mi} is to keep the fields and gauge parameters in $\mbox{ker} \, \Delta$, i.e. the kernel of the second-order differential operator  $\Delta  \equiv 2\partial_{m} \tilde{\partial}^{m}$. Through the projector, some constraints are imposed on momenta of fields and gauge parameters. The weak constraint could be rewritten, in general, in an $O(d, d)$ covariant form:
\bea
\partial_{\tilde{M}}\partial^{\tilde{M}} f=0, 
\eea
where 
\bea
\partial_{\tilde{M}}\equiv 
 \begin{pmatrix} \,\tilde{\partial}^m \, \\[0.6ex] {\partial_m } \end{pmatrix},\qquad m=1, 2, \cdots, d, \qquad \tilde{M}=1, 2,\cdots, 2d.
\eea 
Here, we denote the doubled indices by the uppercase letters and the non-doubled indices by the lower-case letters. The doubled indices are raised or lowered by an $O(d, d)$ invariant metric:
\bea
\eta= \begin{pmatrix} 0& I \\ I& 0 \end{pmatrix}.
\eea
Let us introduce the following notation in the doubled spacetime. We use $X^{\tilde{M}}$  and $K^{\tilde{M}}$ for denoting, respectively:
\begin{eqnarray}
X^{\tilde{M}} \equiv  \left( \begin{array}{c}
                        x^{m}  \\
                        \tilde{x}_{m} \end{array} \right)
                       \,\,\,\,\,\,\,\, \mbox{and} \,\,\,\,\,\,\,\  K^{\tilde{M}} \equiv \left( \begin{array}{c}
                      w^{m}    \\
                       p_{m}  \end{array} \right) 
\end{eqnarray}
where $p_{m}$ are the momenta conjugate to $x^{m}$ while the windings $\omega^{m}$ can be considered as the conjugate to the dual coordinate $\tilde{x}_{m}$.

Before introducing our projector, let us first observe that, given a general double field $A$, one could introduce a Fourier series for it, along the compact dimensions:
\bea
A\equiv\sum_K A_K e^{iKX}
\eea
where $KX\equiv K_{\tilde{M}}X^{\tilde{M}}$.
A canonical projection of the field $A$ into a field satisfying the weak constraint can be defined through the following star product denoted by $*$ \,\, :
\bea
A*1\equiv\sum_K A_K e^{iKX}\delta_{KK,0},
\eea
where $KK\equiv K^{\tilde{M}}K_{\tilde{M}}$. This actually gives
\bea
\partial^{\tilde{M}}\partial_{\tilde{M}}(A*1)=0.
\eea
In other words, the star product imposes the weak constraint on $A$, projecting it into $A*1$ satisfying the constraint $\Delta [A*1] =0$.

It is  also possible to  define the star product of constrained fields $A*1$ and $B*1$ as follows:
\bea
A*B=B*A=\sum_{K, K^{\prime}}A_KB_{K^{\prime}} e^{i (K+K^{\prime} ) X} \delta_{KK, 0} \, \delta_{K^{\prime}K^{\prime}, 0} \, \delta_{KK^{\prime}, 0}.
\eea
Hence, one equivalently obtains:
\bea
\partial^{\tilde{M}}\partial_{\tilde{M}}(A*B)=\partial^{\tilde{M}}A*\partial_{\tilde{M}}B=0.
\eea
This equation shows that actually, through the star product above defined, one could implement the results obtained by imposing the strong constraint in DFT \cite{Hull:2009mi, Siegel:1993xq} as well, starting from fields and gauge parameters only satisfying the weak constraint.
 
One could check the associativity of this operation. Actually, it is straightforward to show that: 
\bea
A*(B*C)&=&\sum_{K, K^{\prime}, K^{\prime\prime}}A_KB_{K^{\prime}}C_{K^{\prime\prime}}
e^{i(K+K^{\prime}+K^{\prime\prime)}X} \, \delta_{KK, 0}\delta_{K^{\prime}K^{\prime}, 0}\delta_{K^{\prime\prime}K^{\prime\prime}, 0}\delta_{K^{\prime}K^{\prime\prime}, 0}
\delta_{K(K^{\prime}+K^{\prime\prime}), 0}\ ,
\nn\\
(A*B)*C&=&\sum_{K, K^{\prime}, K^{\prime\prime}}A_KB_{K^{\prime}}C_{K^{\prime\prime}}
e^{i(K+K^{\prime}+K^{\prime\prime})X}\delta_{KK, 0}\delta_{K^{\prime}K^{\prime}, 0}\delta_{K^{\prime\prime}K^{\prime\prime}, 0}\delta_{KK^{\prime}, 0}
\delta_{(K+K^{\prime})K^{\prime\prime}, 0}\ .
\nn\\
\eea
In general, the star product does not exhibit associativity, but it does under integration since:
\bea
\int dX\ A*(B*C)=\int dX\ (A*B)*C.
\eea
The triple product also satisfies 
\bea
\partial^{\tilde{M}}\partial_{\tilde{M}}\big(A*(B*C)\big)=\partial^{\tilde{M}}\partial_{\tilde{M}}\big((A*B)*C\big)=0.
\eea

 Furthermore, the star product already implies a boundary condition. In fact, it is easy to show that the integral of a total derivative term for any function $f$ is vanishing:
\bea
\int dX\ \partial_{\tilde{M}}f=0  \,\,\,\,\,\,\,\,\,\,   \mbox{with} \,\,\,\,\,\,\,\,\,\, f= \sum_K f_K e^{iKX}
\eea
because $\partial_{\tilde{M}}$ gives a factor proportional to $K_{\tilde{M}}$ and the integration  yields a $\delta_{K,0}$ in the momentum space. The result implies that our fields and gauge parameters need to vanish on the boundary, and have to be periodic if one wants to express them as a Fourier series. DFT makes T-duality manifest in the torus case in which periodic boundary conditions hold, so the projector can be used in this case. 

We stress that using the Fourier expansion for the fields provides a convenient way to understand the weak constraint in terms of vanishing momenta.

\section{$d=1$}
\label{3}
In the case of only one compact dimension, $d=1$, let us say, $x^{m}$, the following constraints on the field momenta 
\bea
KK=0, \qquad K^{\prime}K^{\prime}=0, \qquad K^{\prime\prime}K^{\prime\prime}=0, \qquad (K+K^{\prime})K^{\prime\prime}=0,
\eea
become:
\bea
K=aK^{\prime}=bK^{\prime\prime},
\eea
where $a$ and $b$ are arbitrary constants.
The star product becomes associative because we always need to consider these constraints in the triple products, and the star product in the action can actually be replaced by the ordinary product because of the identity
\bea
&&\int dX\ A\cdot B\cdot C\cdots=\int dX\ A*B*C\cdots\
\eea 
that holds if a boundary term is absent and the Fourier expansion valid.
The identity could be obtained, in this case,  for an arbitrary number of products. 
 The following decomposition of a double field holds: 
\bea
\label{fourier}
f=\sum_{p}f_{p}  e^{ip_{m}x^{m}}+\sum_{\omega}f_{\omega} e^{i\omega^m\tilde{x}_m} 
 \label{fatt}
\eea
since  it has to satisfy $\partial^{\tilde{M}}\partial_{\tilde{M}}f=0$.

Actually,  the strong constraint  removes half of the degrees of freedom, selecting only one of the two functions in \eqref{fourier}. DFT in this case results to be a superposition of the ordinary supergravity and its T-dual. Hence, we do not have any ambiguities in writing the action for $d=1$. This is an interesting point that clarifies what  DFT is.

\section{Arbitrary $d$}
\label{4}
As seen in the previous section, in the case  $d=1$ one can have an associative algebra. 
In the following,  we extend our discussion to arbitrary $d$. 

We consider the triple product first. 
The ordinary triple products have to be accompanied by the conditions:
\bea
KK=0, \qquad K^{\prime}K^{\prime}=0, \qquad K^{\prime\prime}K^{\prime\prime}=0, \qquad K+K^{\prime}+K^{\prime\prime}=0.
\eea
that imply
\bea
K^2=(K^{\prime}+K^{\prime\prime})^2=2K^{\prime}K^{\prime\prime}=0, \qquad K^{\prime 2}=(K^{\prime\prime}+K)^2=2K^{\prime\prime}K=0, 
\nn
\eea
\bea
K^{\prime\prime 2}=(K+K^{\prime})^2=2KK^{\prime}=0  \,\,\, .
\eea
 These constitute orthogonality conditions. The result is interesting because it implies that the cubic action in \cite{Hull:2009mi}, obtained from closed string field theory,  is not modified up to a boundary term when momenta satisfy these conditions that make the star product associative. When the orthogonality condition of momenta is applied, then the closure of the gauge algebra and supersymmetry algebra, and more other properties in strong constrained DFT, can also be obtained. 
 
  We extend this result to an arbitrary number of products of fields with their momenta being orthogonal. Indeed, imposing the orthogonality conditions of momenta beyond the triple products restricts the fluctuations of fields. However, the modification does not give any violation to the known results from closed string field theory \cite{Hull:2009mi}.

The star product still satisfies the identity:
\bea
&&\int dX\ A*B*\cdots=\int dX\ A\cdot B\cdot\cdots .
\eea 
when all momenta are orthogonal to each other if a non-trivial boundary term does not appear. Let us stress here that the orthogonality conditions constitute a way to give associativity to the star product. 

The associativity of the star product could also be obtained by using the weak constraint
\bea
KK=0, \qquad K^{\prime\prime}K^{\prime\prime}=0, \qquad \cdots
\eea
and the vanishing of the sum of the momenta
\bea
K+K^{\prime}+\cdots=0.
\eea
The vanishing of the sum of the momenta would be not a reasonable condition before a boundary term would be omitted. When we consider 
\bea
\int dX\ A
\eea
with the condition of  vanishing momenta, then the fluctuations of  the field  $A$ are lost.

  In order to define a consistent DFT, our approach consists in requiring all momenta to satisfy the orthogonality conditions. This is naturally obtained from the cubic action and guarantees the closure of the symmetry algebra. Such constraints reduce the fluctuations of fields.

\section{Double Field Theory}
\label{5}
We discuss DFT from the generalized metric formulation \cite{Hohm:2010pp} since the action can be uniquely defined from the point of view of the symmetry. The result is the same as in DFT with the strong constraint, up to a boundary term, because the star product is associative. 

\subsection{Generalized Metric Formulation}
We use a general way to introduce the generalized metric formulation from the star product. The action of the generalized metric formulation can be determined by replacing the ordinary product by the star product in the usual generalized metric formulation.

We start from defining the generalized metric ${\cal H}_{MN}$ ($M$, $N$=$1$, $2$, $\cdots$, $2D$):

\bea
  {\cal H}~ \equiv ~ {\cal H}^{\bullet\,\bullet}  \,, \qquad {\cal H} \ \equiv \
  \begin{pmatrix}    g-B*g^{-1}*B & B*g^{-1}\\[0.5ex]
  -g^{-1}*B & g^{-1}\end{pmatrix}\, .
  \eea
  being $g$ the metric tensor, $B$ the Kalb-Ramond field and $d$ the scalar dilaton.

The generalized metric is an $O(D,D)$ symmetric matrix satisfying the relation:
 \be
 {\cal H}*\,\eta\,*{\cal H}=\eta\;.
  \ee
A role of the star product is just to give constraints to the background fields in a doubled spacetime so the generalized metric still has an $O(D, D)$ symmetry.

The inverse of the generalized metric is
\be
  {\cal H}^{-1}~ \equiv ~ {\cal H}_{\bullet\,\bullet}  \,\ = \ \left({\cal H}^{MN}\right)^{-1} \ = \
  \begin{pmatrix}    g^{-1} & -g^{-1}*B\\[0.5ex]
  B*g^{-1} & g-B*g^{-1}*B\end{pmatrix}\;.
  \ee
  and it satisfies the relation:
\bea
{\cal H}^{-1} =  \eta {\cal H} \eta\,.
\eea

 We stress here that ${\cal H}$ and ${\cal H}^{-1}$ are both symmetric matrices. 

In the following, we show how to rewrite the action in terms of the generalized metric and the dilaton. We first assume that this theory has a manifest $O(D, D)$ structure and $\mathbb{Z}_2$  symmetry. The $\mathbb{Z}_2$ symmetry is
\bea
B_{ij}\rightarrow -B_{ij},
\qquad
\tilde{\partial}\rightarrow -\tilde{\partial}.
\eea
This implies that the last transformation 
could be rewritten as
\be
\partial_M\rightarrow  Z\, \partial_M \,, \qquad
Z= \begin{pmatrix} -1 & 0 \\ \phantom{-}0& 1 \end{pmatrix}\,,
\ee
with $Z$ satisfying
\be
Z = Z^T \,,\qquad  Z^2 = 1\,,
\ee
where $T$ denotes the transpose.

Under the transformation $B_{ij} \rightarrow - B_{ij}$, the off-diagonal elements in the generalized metric change sign. This means that
\be
{\cal H}^{MN}  \to  Z {\cal H}^{MN} Z\,, \qquad
{\cal H}_{MN}  \to  Z {\cal H}_{MN} Z \,.
\ee
We also note that $Z$ is not an $O(D,D)$ matrix since
\be
\eta^{MN}  \not=  Z \,\eta ^{MN} Z\,, \qquad
\eta_{MN}  \not=  Z \,\eta_{MN} Z \,.
\ee
By using $\partial_M,  {\cal H}^{MN}$, $ {\cal H}_{MN}$ and $d$, we can construct the gauge invariant action 
by considering all possible terms within second derivative terms up to a boundary term. The action is
\bea
\label{acg}
S_{DFT} &=& \int dx \ d\tilde x  \
   e^{-2d}*\Big(\frac{1}{8}{\cal H}^{MN}*\partial_{M}{\cal H}^{KL}*
  \partial_{N}{\cal H}_{KL}-\frac{1}{2}
  \,{\cal H}^{MN}*\partial_{N}{\cal H}^{KL}*\partial_{L}
  {\cal H}_{MK}
\nn\\
  &&-2\partial_{M}d*\partial_{N}{\cal H}^{MN}+4{\cal H}^{MN}*\,\partial_{M}d
  *\partial_{N}d \Big),
 \eea
 where
 \bea
   e^{-2d}\equiv\sqrt{-g}*e^{-2\phi},
  \eea
 The gauge transformations are provided by:
\bea
\delta_{\xi} d&=&-\frac{1}{2}\partial_M\xi^M+\xi^M*\partial_M d,
\nn\\
\delta_{\xi} {\cal H}^{MN}&=&\hat{{\cal L}}_{\xi}{\cal H}^{MN}\equiv\xi^P*\partial_P{\cal H}^{MN}+(\partial^M\xi_p-\partial_P\xi^M)*{\cal H}^{PN}+(\partial^N\xi_P-\partial_P\xi^N)*{\cal H}^{MP}.
\nn\\
\eea
and exhibit a closure property:
\bea
[\delta_{\xi_1}, \delta_{\xi_2}]=-\delta_{[\xi_1, \xi_2]_C},
\eea
where the $C$-bracket is defined by
\bea
[\xi_1, \xi_2]_C^M=\xi_1^N*\partial_N\xi_2^M-\xi_2^N*\partial_N\xi_1^M-\frac{1}{2}\eta^{MN}\eta_{PQ}\xi_1^P*\partial_N\xi_2^Q+\frac{1}{2}\eta^{MN}\eta_{PQ}\xi_2^P*\partial_N\xi_1^Q.
\nn\\
\eea

If we assume that all fields in the action are independent of the $\tilde{x}$'s or their dual coordinates, it can be easily checked that the action would reduce to the ordinary supergravity theory
\bea
\int dx\ \sqrt{-g}e^{-2\phi}\bigg( R+4(\partial\phi)^2-\frac{1}{12}H^2\bigg),
\eea
where $\phi$ is the dilaton satisfying $\sqrt{-g}e^{-2\phi}=e^{-2d}$, $R$ is the Ricci scalar and $H=dB$ is the three-form field strength. 
The closure of the gauge algebra also guarantees the closure of the supersymmetry algebra because the star product is associative.

Since the star product is associative, the background independence to all orders at classical level is also included in this action \cite{Hohm:2015ugy}. When we consider $d=1$, DFT is also useful for studying the non-geometric $R$-flux from a constant $H$-flux \cite{Andriot:2011uh}.

Finally, we summarize the assumptions that we have used to find the proposed action. We have first used the $O(D, D)$ symmetry structure and $\mathbb{Z}_2$ symmetry to find all possible terms for second derivative terms up to a boundary term. We also expect that this theory could reduce to the ordinary supergravity so all terms in the action should couple to $e^{-2d}$. Finally, we have used the gauge symmetry to find the coefficients up to an overall constant. The overall constant are determined by requiring that DFT  to reduce to the ordinary supergravity when all fields are independent of the dual coordinates.

\section{Double Sigma Model}
\label{6}
We propose a double sigma model already discussed in \cite{Copland:2011wx, Park:2016sbw} by including the star product, establishing their equivalence by showing that one could get the same equations of motion. Moreover,  an interpretation of  the star product in the world-sheet space is provided in the simplest case $d=1$ .

\subsection{Action}
The double sigma model is

\bea
S_{\Sigma}=\frac{1}{2}\int d^2\sigma\ \bigg(\partial_1X^A*{\cal H}_{AB}*\partial_1X^B-\partial_1X^A*\eta_{AB}*\partial_0X^B\bigg),
\eea
Since one-loop $\beta$ functions \cite{Copland:2011wx} could reproduce the equations of motion of the DFT without using any constraints, we propose that the star product should also appear in the double sigma model to reproduce the equations of motion of the DFT defined by the star product.

\subsection{Classical Equivalence}
In this section, the classical equivalence between the double sigma model and ordinary sigma model with an on-shell self-duality relation for $d=1$ is shown. The result implies that we could find the same equations of motion as in the case of the ordinary sigma model. The equations of motion deriving from the double sigma  model are:
\bea
\label{beom}
\partial_1\bigg({\cal H}_{mA}*\partial_1X^A-\eta_{mA}\partial_0X^A\bigg)&=&\frac{1}{2}\partial_1X^A*\partial_m{\cal H}_{AB}*\partial_1X^B,
\nn\\
\partial_1\bigg({\cal H}^m{}_A*\partial_1X^A-\eta^m{}_A\partial_0X^A\bigg)&=&\frac{1}{2}\partial_1X^A*\tilde{\partial}^m{\cal H}_{AB}*\partial_1X^B.
\eea
When we consider $d=1$, the background fields can be decomposed into the sum of the ordinary background and dual background fields.  First considering ordinary background fields,  we have:
\bea
\partial_1\bigg({\cal H}^m{}_A\partial_1X^A-\eta^m{}_A\partial_0X^A\bigg)=0.
\eea
A suitable self-duality relation for the case of the ordinary background fields is
\bea
{\cal H}^m{}_A\partial_1X^A-\eta^m{}_A\partial_0X^A=0.
\eea
The self-duality relation is also equivalent to
\bea
\partial_1\tilde{X}_m=g_{mn}\partial_0X^n+B_{mn}\partial_1X^n.
\eea
The other equation of motion is
\bea
&&\partial_1\bigg\lbrack\bigg(g-Bg^{-1}B\bigg)_{mn}\partial_1X^n+\bigg(Bg^{-1}\bigg)_m{}^n\partial_1\tilde{X}_n-\partial_0\tilde{X}_m\bigg\rbrack
\nn\\
&=&\frac{1}{2}\partial_1X^p\partial_m\bigg(g-Bg^{-1}B\bigg)_{pq}\partial_1X^q+\partial_1X^p\partial_m\bigg(Bg^{-1}\bigg)_p{}^q\partial_1\tilde{X}_q+\frac{1}{2}\partial_1\tilde{X}_p\partial_mg^{pq}\partial_1\tilde{X}_q
\nn\\
\eea
when we consider the case of the ordinary background fields.

We find the same equations of motion as in the ordinary sigma model by using the self-duality relation to remove the dual coordinates as
\bea
&&\partial_1\bigg\lbrack\bigg(g-Bg^{-1}B\bigg)_{mn}\partial_1X^n+\bigg(Bg^{-1}\bigg)_m{}^n\partial_1\tilde{X}_n-\partial_0\tilde{X}_m\bigg\rbrack
\nn\\
&=&\partial_1\bigg(g_{mn}\partial_1X^n+B_{mn}\partial_0X^n\bigg)-\partial_0\bigg(g_{mn}\partial_0X^n+B_{mn}\partial_1X^n\bigg) \,\,\, ;
\eea
\bea
&&\frac{1}{2}\partial_1X^p\partial_m\bigg(g-Bg^{-1}B\bigg)_{pq}\partial_1X^q+\partial_1X^p\partial_m\bigg(Bg^{-1}\bigg)_p{}^q\partial_1\tilde{X}_q+\frac{1}{2}\partial_1\tilde{X}_p\partial_mg^{pq}\partial_1\tilde{X}_q
\nn\\
&=&-\frac{1}{2}\partial_0X^p\partial_mg_{pq}\partial_0X^q+\frac{1}{2}\partial_1X^p\partial_mg_{pq}\partial_1X^q+\partial_1X^p\partial_mB_{pq}\partial_0X^q.
\eea

Then we consider the dual background fields and we get
\bea
\partial_1\bigg({\cal H}_{mA}\partial_1X^A-\eta_{m_A}\partial_0X^A\bigg)=0.
\eea
We consider the self-duality relation:
\bea
{\cal H}_{mA}\partial_1X^A-\eta_{mA}\partial_0X^A=0.
\eea
and it is equivalent to
\bea
\partial_0\tilde{X}_m=-G_{mn}\beta^{np}\partial_1\tilde{X}_p+G_{mn}\partial_1X^n
\eea
where the dual fields $G$ and $\beta$ appear.

The other equation of motion is
\bea
&&\partial_1\bigg\lbrack\bigg(G^{-1}-\beta G\beta\bigg)^{mn}\partial_1\tilde{X}_n+\bigg(\beta G\bigg)^m{}_n\partial_1 X^n-\partial_0 X^m\bigg\rbrack
\nn\\
&=&\frac{1}{2}\partial_1X^p\partial^m G_{pq}\partial_1X^q-\partial_1X^p\partial^m\bigg(G\beta\bigg)_p{}^q\partial_1\tilde{X}_q+\frac{1}{2}\partial_1\tilde{X}_p\partial^m\bigg(G^{-1}-\beta G\beta\bigg)^{pq}\partial_1\tilde{X}_q.
\nn\\
\eea
We find the same equations of motion as in the dual sigma model by using the self-duality relation to remove the dual coordinates as
\bea
&&\partial_1\bigg\lbrack\bigg(G^{-1}-\beta G\beta\bigg)^{mn}\partial_1\tilde{X}_n+\bigg(\beta G\bigg)^m{}_n\partial_1 X^n-\partial_0 X^m\bigg\rbrack
\nn\\
&=&\partial_1\bigg(G^{mn}\partial_1\tilde{X}_n+\beta^{mn}\partial_0\tilde{X}_n\bigg)-\partial_0\bigg(G^{mn}\partial_0\tilde{X}_n+\beta^{mn}\partial_1\tilde{X}_n\bigg) ;
\eea
\bea
&&\frac{1}{2}\partial_1X^p\partial^m G_{pq}\partial_1X^q-\partial_1X^p\partial^m\bigg(G\beta\bigg)_p{}^q\partial_1\tilde{X}_q+\frac{1}{2}\partial_1\tilde{X}_p\partial^m\bigg(G^{-1}-\beta G\beta\bigg)^{pq}\partial_1\tilde{X}_q
\nn\\
&=&-\frac{1}{2}\partial_0\tilde{X}_p\partial^mG^{pq}\partial_0\tilde{X}_q+\frac{1}{2}\partial_1\tilde{X}_p\partial^mG^{pq}\partial_1\tilde{X}_q+\partial_1\tilde{X}_p\partial^m\beta^{pq}\partial_0\tilde{X}_q.
\eea

The equations of motion of the double sigma model are then shown to be the equations  of motion of the ordinary and dual sigma model from the on-shell self-duality relation. Thus, we obtain the consistent result for $d=1$.
\subsection{Off-Shell Self-Duality Relation}
We implement the self-duality relation at the off-shell level to obtain the equivalence of the equations of motion. The equations of motion for the case of the ordinary background fields in the bulk are
\bea
&&\partial_1\bigg(g^{-1}\partial_1\tilde{X}-g^{-1}B\partial_1X-\partial_0X\bigg)^m=0,
\nn\\
&&\partial_1\bigg(Bg^{-1}\partial_1\tilde{X}+\big(g-Bg^{-1}B\big)\partial_1X-\partial_0\tilde{X}\bigg)_m
\nn\\
&=&\frac{1}{2}\partial_1X\partial_m\bigg(g-Bg^{-1}B\bigg)\partial_1X+\partial_1X\partial_m\bigg(Bg^{-1}\bigg)\partial_1\tilde{X}+\frac{1}{2}\partial_1\tilde{X}\partial_mg^{-1}\partial_1\tilde{X}.
\eea
We could shift $X^m$ ($X^m\rightarrow X^m+f^m(\sigma^0)$) and redefine $g$ and $B$ to obtain
\bea
&&\partial_1\tilde{X}_m=B_{mn}\partial_1X^n+g_{mn}\partial_0X^n,
\nn\\
&&\partial_1\bigg(g_{mn}\partial_1X^n+B_{mn}\partial_0X^n\bigg)-\partial_0\bigg(g_{mn}\partial_0X^n+B_{mn}\partial_1X^n\bigg)
\nn\\
&=&-\frac{1}{2}\partial_0X^p\partial_mg_{pq}\partial_0X^q+\frac{1}{2}\partial_1X^p\partial_mg_{pq}\partial_1X^q+\partial_1X^p\partial_mB_{pq}\partial_0X^q.
\eea
When we consider the dual background fields, the equations of motion in the bulk are
\bea
&&\partial_1\bigg(-G\beta\partial_1\tilde{X}+G\partial_1X-\partial_0\tilde{X}\bigg)_m=0,
\nn\\
&&\partial_1\bigg(\big(G^{-1}-\beta G\beta\big)\partial_1\tilde{X}+\big(\beta G\big)\partial_1X-\partial_0 X\bigg)^m
\nn\\
&=&\frac{1}{2}\partial_1X\partial^mG\partial_1X-\partial_1X\partial^m\bigg(G\beta\bigg)\partial_1\tilde{X}+\frac{1}{2}\partial_1\tilde{X}\partial_m\bigg(G^{-1}-\beta G\beta\bigg)\partial_1\tilde{X}.
\eea
We could shift $\tilde{X}_m$ ($\tilde{X}_m\rightarrow \tilde{X}_m+f_m(\sigma^0)$) and redefine $G$ and $\beta$. Then we obtain
\bea
&&\partial_0\tilde{X}_m=-G_{mn}\beta^{np}\partial_1\tilde{X}_p+G_{mn}\partial_1X^n,
\nn\\
&&\partial_1\bigg(G^{mn}\partial_1\tilde{X}_n+\beta^{mn}\partial_0\tilde{X}_n\bigg)-\partial_0\bigg(G^{mn}\partial_0\tilde{X}_n+\beta^{mn}\partial_1\tilde{X}_n\bigg)
\nn\\
&=&-\frac{1}{2}\partial_0\tilde{X}_p\partial^mG^{pq}\partial_0\tilde{X}_q+\frac{1}{2}\partial_1\tilde{X}_p\partial^mG^{pq}\partial_1\tilde{X}_q+\partial_1\tilde{X}_p\partial^m\beta^{pq}\partial_0\tilde{X}_q.
\eea

Thus, we could get the equations of motion of the ordinary and the dual sigma model from the double sigma model for $d=1$ from the self-dual relations at the off-shell level. This gives the consistency of the double sigma model with the modification of the star product. When we use the one-loop $\beta$-functions to obtain the equations of motion of the DFT \cite{Copland:2011wx}, the equations of motion should be modified by the star product when a non-trivial boundary term appears. The additional of the star product in the double sigma model should give us correct equations of motion of the DFT.

\subsection{ Interpretation of the Star Product in the World-Sheet }
The star product in the world-sheet space is not so clear as in the target space. For $d=1$, DFT should be the sum of the ordinary supergravity and the dual supergravity. Hence, we expect that the double sigma model should be the sum of the ordinary sigma model and the dual sigma model. The unclear part of the star product in the worldsheet space is the Fourier expansion in the target space. When we consider the first term of the double sigma model
\bea
\frac{1}{2}\int d^2\sigma\ \partial_1X^A{\cal H}_{AB}\partial_1 X^B,
\eea
we expect that the dual target space should be auxiliary when the generalized metric only depends on the ordinary target space. The star product for $d=1$ should play the same kind of role in the world-sheet as in the target space. Thus, the partition function of the double sigma model for $d=1$ can be rewritten as
\bea
\int DX\exp\bigg\lbrack \frac{1}{2}\int d^2\sigma\ \bigg(\partial_1X^A{\cal H}_{AB}(X^m)\partial_1X^B-\partial_1X^A\eta_{AB}\partial_0X^B\bigg)\bigg\rbrack
\nn\\
\times\int DX^{\prime}\exp\bigg\lbrack \frac{1}{2}\int d^2\sigma\ \bigg(\partial_1X^{\prime A}{\cal H}_{AB}(\tilde{X}^{\prime}_m)\partial_1X^{\prime B}-\partial_1X^{\prime A}\eta_{AB}\partial_0X^{\prime B}\bigg)\bigg\rbrack .
\eea
When we integrate $\tilde{X}_m$ out, we can obtain the ordinary sigma model and we only have the dynamical terms of $X^m$. If we integrate out $X^{\prime m}$, then the dual sigma model is obtained and its dynamics is determined by $\tilde{X}_m^{\prime}$. Although  the star product seems to double the target space, the physical degrees of freedom of the target space are not increased by using the star product. 

\section{T-Duality in a Non-Commutative Space}
\label{7}
We discuss T-duality in a non-commutative space \cite{Kamani:2001yd} and we use $O(D, D)$ transformations to connect T-duality and the relation between the closed string parameters $g$ and $B$ and the open string parameters $G$, $\Pi$ and $\Phi$ \cite{Jurco:2013upa}. The latter can be defined through the following relation: 
\bea
 \begin{pmatrix} g & B \\ B & g \end{pmatrix}^{-1}=\begin{pmatrix} G & \Phi
 \\ -\Phi^T & G \end{pmatrix}^{-1}+ \begin{pmatrix} 0 & \Pi
 \\ -\Pi^T & 0 \end{pmatrix}
\eea
This can be understood from the point of view of $O(D, D)$ transformations \cite{Jurco:2013upa} as follows:
\bea
&&\begin{pmatrix} 1 & \Pi
 \\ 0 & 1 \end{pmatrix}*
\begin{pmatrix} 1 & 0
 \\ -\Phi^T & 1 \end{pmatrix}*
\begin{pmatrix} G^{-1} & 0
 \\ 0 & G \end{pmatrix}*
\begin{pmatrix} 1 & -\Phi
 \\ 0 & 1 \end{pmatrix}*
\begin{pmatrix} 1 & 0
 \\ \Pi^T & 1 \end{pmatrix}
 \nn\\
 &=&\begin{pmatrix} (1-\Pi*\Phi^T)*G^{-1}*(1-\Phi*\Pi^T) +\Pi* G*\Pi^T & -(1-\Pi*\Phi^T)*G^{-1}*\Phi+\Pi *G
 \\ -\Phi^T*G^{-1}*(1-\Phi*\Pi^{T})+G*\Pi^T & \Phi^T*G^{-1}*\Phi+G \end{pmatrix}
 \nn\\
 &=&\begin{pmatrix} g^{-1} & -g^{-1}*B
 \\ -B^T*g^{-1} & g+B^T*g^{-1}*B \end{pmatrix}.
\eea
Let us consider here  the open-closed string parameters relation
\bea
\frac{1}{g+B}=\frac{1}{G+\Phi}+\Pi,
\eea
which also contains T-duality in the commutative space, and T-duality in the non-commutative space
\bea
\label{nct}
\frac{1}{\tilde{G}+\tilde{\Phi}}=(g-B)*(G-\Phi)^{-1}*(g+B), \qquad \tilde{\Pi}=-(g-B)*\Pi*(g+B).
\eea
Here, the $\tilde{G}$ and $\tilde{\Phi}$ are the dual open string parameters, the generalized metric in the commutative space
\bea
\label{gcs}
\begin{pmatrix} g^{-1} & -g^{-1}*B
 \\ -B^T*g^{-1} & g+B^T*g^{-1}*B \end{pmatrix}
\eea
and the generalized metric in the non-commutative space
\bea
\begin{pmatrix} (1-\Pi*\Phi^T)*G^{-1}*(1-\Phi*\Pi^T)+\Pi* G*\Pi^T & -(1-\Pi*\Phi^T)*G^{-1}*\Phi+\Pi *G
 \\ -\Phi^T*G^{-1}*(1-\Phi*\Pi^{T})+G*\Pi^T & \Phi^T*G^{-1}*\Phi+G \end{pmatrix}
 \nn\\
\eea
One can then use \eqref{nct} in order 
to obtain the dual generalized metric in the commutative space 
\bea
\begin{pmatrix} \tilde{g}+\tilde{B}*\tilde{g}^{-1}*\tilde{B}^T & \tilde{B}*\tilde{g}^{-1}
 \\ \tilde{g}^{-1}*\tilde{B}^T & \tilde{g}^{-1} \end{pmatrix}.
\eea
in which the $\tilde{g}$ and $\tilde{B}$ are the dual closed string parameters,
and the dual generalized metric in the non-commutative space
\bea
\begin{pmatrix} \tilde{\Phi}^T*\tilde{G}^{-1}*\tilde{\Phi}+\tilde{G} &-\tilde{\Phi}^T*\tilde{G}^{-1}*(1-\tilde{\Phi}*\tilde{\Pi}^{T})+\tilde{G}*\tilde{\Pi}^T
 \\  -(1-\tilde{\Pi}*\tilde{\Phi}^T)*\tilde{G}^{-1}*\tilde{\Phi}+\tilde{\Pi}*\tilde{G} & (1-\tilde{\Pi}*\tilde{\Phi}^T)*\tilde{G}^{-1}*(1-\tilde{\Phi}*\tilde{\Pi}^T)+\tilde{\Pi}* \tilde{G}*\tilde{\Pi}^T \end{pmatrix}.
\nn\\
\eea

Our conclusion is that the $O(D, D)$ transformations could give us the generalized metric and the dual generalized metric in the commutative space and the non-commutative space.

\section{Open String Theory for One Doubled Space}
We construct the Neumann and Dirichlet boundary conditions from the equivalence of the equations of motion by the self-duality relation \cite{Ma:2015yma}. When we only double one space coordinate, the partition function of the double sigma model should be the multiplication of the partition function of the ordinary sigma model and the partition function of the dual sigma model. We also obtain this consistent answer when we include the Neumann and Dirichlet boundary conditions in the double sigma model. Finally, we exhibit the low-energy effective theory for the open string theory and consider the non-commutative geometry through the $O(D, D)$ transformations \cite{Ma:2014vqm}.
\label{8}
\subsection{Classical Equivalence}
We include the open string theory from the classical equivalence by the on-shell self-duality relation already given for $d=1$.
The equations of motion of the double sigma model in the bulk are in \eqref{beom}.

In the ordinary sigma model, the well-known Neumann boundary conditions are
\bea
g_{mn}\partial_1X^n+B_{mn}\partial_0X^n=0.
\eea
In order to obtain the Neumann boundary conditions, we impose 
\bea
{\cal H}_{mA}\partial_1X^A=0
\eea
on the $\sigma^1$ direction. These boundary conditions can be rewritten as:
\bea
Bg^{-1}\partial_1\tilde{X}+(g-Bg^{-1}B)\partial_1X=0.
\eea
We use the following relations:
\bea
{\cal H}^m{}_A\partial_1X^A-\eta^m{}_A\partial_0X^A=0, \qquad {\cal H}_{mA}\partial_1X^A=0
\eea
to obtain the Neumann boundary conditions in the ordinary sigma model. 

Let us consider the dual background and  choose the boundary conditions on the $\sigma^1$ direction as:
\bea
{\cal H}_{mA}\partial_1X^A-\eta_{mA}\partial_0X^A=0, \qquad {\cal H}^m{}_A\partial_1X^A=0.
\eea
These can be rewritten as
\bea
G\partial_1X-G\beta\partial_1\tilde{X}-\partial_0\tilde{X}=0, \qquad (G^{-1}-\beta G\beta)\partial_1\tilde{X}+\beta G\partial_1X=0
\eea
by using the $G$ and $\beta$. Therefore, we also obtain the Neumann boundary conditions on the $\sigma^1$ direction 
\bea
G^{-1}\partial_1\tilde{X}+\beta\partial_0\tilde{X}=0
\eea
in the dual sigma model. Combining the boundary conditions that we used on the $\sigma_1$ direction in the ordinary and dual sigma models, then the boundary conditions can be used in a simple way since they can be expressed as follows:
\bea
\bigg({\cal H}\partial_1X-\eta\partial_0X\bigg)^M=0\,\,\,\,\,\,\,\,\,\, , \qquad \bigg({\cal H}\partial_1X\bigg)_M=0.
\eea
This implies that an $O(D, D)$ boundary condition is necessary when the star product is used in the double sigma model. We can have different $O(D, D)$ solutions through the $O(D, D)$ transformations. Each background solution has a corresponding boundary condition to let the string end on the brane. 

We list here the boundary conditions in the double sigma model 
\bea
\bigg({\cal H}\partial_1X-\eta\partial_0X\bigg)^M=0, \qquad \bigg({\cal H}\partial_1X\bigg)_M=0, \qquad \bigg(\eta\partial_0X\bigg)_M=0
\eea
on the $\sigma^1$ direction and 
\bea
(\delta X)^M=0
\eea
on the $\sigma^0$ direction. 

The equivalence of the equations of motion in the double sigma model with boundary conditions at the off-shell level should be directly related to whether we could obtain the self-dual relation at this level. When we consider closed string theory in the double sigma model, we could use the off-shell way to obtain the self-dual relation. Thus, the equivalence could be extended to the off-shell level even for the double sigma model with the boundary conditions.

\subsection{Double Sigma Model with Boundary Conditions}
We consider the double sigma model with the boundary conditions, then we can obtain the ordinary and dual sigma models. For one doubled space, the partition function of the double sigma model with the boundary conditions is
\bea
\int DX\exp\bigg\lbrack \frac{1}{2}\int d^2\sigma\ \bigg(\partial_1X^A{\cal H}_{AB}(X^m)\partial_1X^B-\partial_1X^A\eta_{AB}\partial_0X^B\bigg)\bigg\rbrack
\nn\\
\times\int DX^{\prime}\exp\bigg\lbrack \frac{1}{2}\int d^2\sigma\ \bigg(\partial_1X^{\prime A}{\cal H}_{AB}(\tilde{X}^{\prime}_m)\partial_1X^{\prime B}-\partial_1X^{\prime A}\eta_{AB}\partial_0X^{\prime B}\bigg)\bigg\rbrack .
\eea
If we first integrate $X^{m}$, we can obtain the ordinary sigma model. When we perform the integration, the result is equivalent to using
\bea
\partial_1\tilde{X}_p=g_{pn}\partial_0X^n+B_{pn}\partial_1X^n
\eea
in the exponent.
Then the result of integration in the exponent is
\bea
&&
\frac{1}{2}\partial_1X^m\bigg(g-Bg^{-1}B\bigg)_{mn}\partial_1X^n
+\partial_1X^m\bigg(Bg^{-1}\bigg)_m{}^n\partial_1\tilde{X}_n
+\frac{1}{2}\partial_1\tilde{X}_m\bigg(g^{-1}\bigg)^{mn}\partial_1\tilde{X}^n
\nn\\
&&-\partial_1\tilde{X}_m\partial_0X^m
\nn\\
&=&\frac{1}{2}\partial_1X^m\bigg(g-Bg^{-1}B\bigg)_{mn}\partial_1X^n
+\frac{1}{2}\partial_1X^m\bigg(Bg^{-1}\bigg)_m{}^n\partial_1\tilde{X}_n
-\frac{1}{2}\partial_1\tilde{X}_m\partial_0X^m
\nn\\
&=&
\frac{1}{2}\partial_1X^m\bigg(g-Bg^{-1}B\bigg)_{mn}\partial_1X^n
+\frac{1}{2}\partial_1X^mB_{mn}\partial_0X^n+\frac{1}{2}\partial_1X^m\bigg(Bg^{-1}B\bigg)_{mn}\partial_1X^n
\nn\\
&&
-\frac{1}{2}\partial_0X^mg_{mn}\partial_0X^n
+\frac{1}{2}\partial_1X^mB_{mn}\partial_0X^n
\nn\\
&=&-\frac{1}{2}\partial_0X^mg_{mn}\partial_0X^n+\frac{1}{2}\partial_1X^mg_{mn}\partial_1X^n+\partial_1X^mB_{mn}\partial_0X^n.
\eea
When we integrate $X^m$ out, we also have a determinant term in the measure that is given by:
\bea
\int DX^m\sqrt{\det g}.
\eea

By  integrating $X^{\prime m}$ out  we obtain the dual sigma model. The integration gives
\bea
\partial_0\tilde{X}^{\prime}_p=G_{pn}\partial_1X^{\prime n}-G_{pm}\beta^{mn}\partial_1\tilde{X}^{\prime n}
\eea
in the exponent. The result of integration in the exponent can be shown to be rewritten as:
\bea
&&
\frac{1}{2}\partial_1X^{\prime m}G_{mn}\partial_1X^{\prime n}
-\partial_1X^{\prime m}\bigg(G\beta\bigg)_m{}^n\partial_1\tilde{X}^{\prime}_n
+\frac{1}{2}\partial_1\tilde{X}^{\prime}_m\bigg(G^{-1}-\beta G\beta\bigg)^{mn}\partial_1\tilde{X}^{\prime n}
\nn\\
&&-\partial_0\tilde{X}^{\prime}_m\partial_1X^{\prime m}
\nn\\
&=&\frac{1}{2}\bigg(\partial_0\tilde{X}^{\prime}G^{-1}-\partial_1\tilde{X}^{\prime}\beta\bigg)G\bigg(G^{-1}\partial_0\tilde{X}^{\prime}+\beta\partial_1\tilde{X}^{\prime}\bigg)-\bigg(\partial_0\tilde{X}^{\prime}G^{-1}-\partial_1\tilde{X}^{\prime}\beta\bigg)G\beta\partial_1\tilde{X}^{\prime}
\nn\\
&&+\frac{1}{2}\partial_1\tilde{X}^{\prime}\bigg(G^{-1}-\beta G\beta\bigg)\partial_1\tilde{X}^{\prime}-\partial_0\tilde{X}^{\prime}\bigg(G^{-1}\partial_0\tilde{X}^{\prime}+\beta\partial_1\tilde{X}^{\prime}\bigg)
\nn\\
&=&\frac{1}{2}\partial_0\tilde{X}^{\prime}G^{-1}\partial_0\tilde{X}^{\prime}+\frac{1}{2}\partial_0\tilde{X}^{\prime}\beta\partial_1\tilde{X}^{\prime}-\frac{1}{2}\partial_1\tilde{X}^{\prime}\beta\partial_0\tilde{X}^{\prime}-\frac{1}{2}\partial_1\tilde{X}^{\prime}\beta G\beta\partial_1\tilde{X}^{\prime}-\partial_0\tilde{X}^{\prime}\beta\partial_1\tilde{X}^{\prime}
\nn\\
&&
+\partial_1\tilde{X}^{\prime}\beta G\beta\partial_1\tilde{X}^{\prime}
+\frac{1}{2}\partial_1\tilde{X}^{\prime}G^{-1}\partial_1\tilde{X}^{\prime}-\frac{1}{2}\partial_1\tilde{X}^{\prime}\beta G\beta\partial_1\tilde{X}^{\prime}-\partial_0\tilde{X}^{\prime}G^{-1}\partial_0\tilde{X}^{\prime}-\partial_0\tilde{X}^{\prime}\beta\partial_1\tilde{X}^{\prime}
\nn\\
&=&-\frac{1}{2}\partial_0\tilde{X}^{\prime}G^{-1}\partial_0\tilde{X}^{\prime}+\frac{1}{2}\partial_1\tilde{X}^{\prime}G^{-1}\partial_1\tilde{X}^{\prime}+\partial_1\tilde{X}^{\prime}\beta\partial_0\tilde{X}^{\prime}.
\eea
We also get the following measure
\bea
\int D\tilde{X}^{\prime}_m\sqrt{\det G^{-1}}
\eea
after integrating $X^{\prime m}$ out.

At this point, we can obtain the ordinary and dual sigma model after doing the integration from the double sigma models with the boundary conditions. An interesting property is that the measure in the double sigma model is still invariant under diffeomorphisms  with a shift symmetry. Thus, we show that the $O(D, D)$ measure still contains the local symmetry although the $O(D, D)$ structure is a global structure. The local symmetry in the $O(D, D)$ structure is defined by summing over all different sectors of DFT.

\subsection{The Low-Energy Effective Theory}
The low-energy effective theory  can  be obtained through one-loop $\beta$-functions \cite{Abouelsaood:1986gd}. The low-energy effective theory for the open string theory from the double sigma model for only one doubled space dimension is the sum of the ordinary DBI action and the dual DBI action. The theory can be written as
\bea
\int dX\ e^{-d}*\bigg(-\det{} {\cal H}_{mn}\bigg)^{\frac{1}{4}}.
\eea

The theory can actually be rewritten in terms of the closed string parameters instead of the open string ones to show the equivalence between the non-commutative and commutative descriptions \cite{Jurco:2013upa} in the DBI theory through the star product because the form of action is not changed after the introduction of the star product. 

The physical picture of the ordinary sigma model for open strings consists essentially in seeing at them ending on a brane. The physical picture of the double sigma model with boundary condition is enlarged via the star product or $O(D, D)$ transformation. For only one doubled spatial dimension, each field could decompose into the ordinary field and dual field. In other words, one choice is the ordinary spacetime, and the other choice is the dual spacetime. These two choices have corresponding boundary conditions to let string end on the brane, because different sectors of DFT are independent. Thus, the boundary conditions of different sectors of DFT  should be independent. In order to double more dimensions, the discussion should be similar.

\section{Conclusion}
\label{9}
We have constructed a DFT-inspired low-energy effective action. A difficulty in DFT is that the closure of the gauge algebra is lost even if the weak constraint, coming from closed string theory, is imposed \cite{Hull:2009mi}. In order to solve this problem, we have analyzed the simplest case of only one doubled space coordinate, then the non-associativity of the star product disappears. Thus, this outcome directly exhibits the closure of the symmetry transformations. Especially for the interesting case $d=1$, the action up to a boundary term can be determined uniquely by the gauge symmetry and the weak constraint can imply the orthogonality of momenta. Thus, this provides a very concrete example to understand DFT. For giving a more generic understanding to the closure of the symmetry, we have considered the triple products for arbitrary $d$, so finding that the triple products defined in the ordinary product could give the orthogonality conditions of momenta. When this property is extended to an arbitrary number of products, one can see the equivalence between the star product and the ordinary product if a non-trivial boundary term is absent. Thus, this theory contains the Einstein gravity theory, closed gauge algebra, supersymmetry algebra and background independence. These properties should be the minimum requirement of a supergravity theory. An interesting property in the ordinary supergravity is that this could be related to the ordinary sigma model via the one-loop $\beta$-functions and the method should provide a background independent theory. Based on the result of the one-loop $\beta$-functions in the double sigma model \cite{Copland:2011wx}, we have used the star product to define the double sigma model. The evidence of the star product in the double sigma model can also be obtained from the classical equivalence by the self-duality relation. We have also found that the $O(D, D)$ transformations can build the relations between the closed string and open string parameters and the T-duality transformation can be obtained from the $O(D, D)$ transformations in DFT. The double sigma model with the Neumann boundary conditions and the Dirichlet boundary conditions has also been constructed. We have also considered the classical equivalence, the low-energy effective theory, and the non-commutative geometry for one doubled space dimension in the context of the open string theory.

Closed string field theory gives the cubic action up to a boundary term and this is what is required to DFT to reproduce. The orthogonality imposed on the momenta does not modify the result of the cubic action up to a boundary term so it shows that all physical observables do not change. The modification of the gauge symmetry with the on-shell constraint does not modify the physical results in the cubic action up to a boundary term. Thus, our proposal could be seen as finding a loophole of the results of closed string field theory \cite{Hull:2009mi}. To find more evidences for the proposal, we need to consider computations of higher order terms from closed string field theory. 

\section*{Acknowledgments}
We would like to thank Martin Cederwall, Olaf Hohm, Kanghoon Lee, Jeong-Hyuck Park and Masaki Shigemori for useful discussions. Especially, Chen-Te Ma would like to thank Nan-Peng Ma for his suggestion and encouragement.

Moreover, both of the authors would like to acknowledge the hospitality of the APCTP  in the Pohang (Korea) during the workshop {\em Duality and Novel Geometry in M-Theory}. F.P. also thanks, for the same reason, the Simons Center during the Summer Workshop in Mathematics and Physics 2016 and the Galileo Galilei Institute in Florence during the workshop {\em Supergravity: what next?}.



\baselineskip 22pt
\begin{thebibliography}{99}
\bibitem{Abouelsaood:1986gd} 
  A.~Abouelsaood, C.~G.~Callan, Jr., C.~R.~Nappi and S.~A.~Yost,
  ``Open Strings in Background Gauge Fields,''
  Nucl.\ Phys.\ B {\bf 280}, 599 (1987).
  doi:10.1016/0550-3213(87)90164-7
  C.~G.~Callan, Jr., C.~Lovelace, C.~R.~Nappi and S.~A.~Yost,
  ``String Loop Corrections to beta Functions,''
  Nucl.\ Phys.\ B {\bf 288}, 525 (1987).
  doi:10.1016/0550-3213(87)90227-6
  
\bibitem{Siegel:1993xq} 
  W.~Siegel,
  ``Two vierbein formalism for string inspired axionic gravity,''
  Phys.\ Rev.\ D {\bf 47}, 5453 (1993)
  doi:10.1103/PhysRevD.47.5453
  [hep-th/9302036].
  W.~Siegel,
  ``Superspace duality in low-energy superstrings,''
  Phys.\ Rev.\ D {\bf 48}, 2826 (1993)
  doi:10.1103/PhysRevD.48.2826
  [hep-th/9305073].
  W.~Siegel,
  ``Manifest duality in low-energy superstrings,''
  hep-th/9308133.

\bibitem{Hull:2009mi} 
  C.~Hull and B.~Zwiebach,
  ``Double Field Theory,''
  JHEP {\bf 0909}, 099 (2009)
  doi:10.1088/1126-6708/2009/09/099
  [arXiv:0904.4664 [hep-th]].

\bibitem{Hohm:2010jy} 
  O.~Hohm, C.~Hull and B.~Zwiebach,
  ``Background independent action for double field theory,''
  JHEP {\bf 1007}, 016 (2010)
  doi:10.1007/JHEP07(2010)016
  [arXiv:1003.5027 [hep-th]].

\bibitem{Hohm:2015ugy} 
  O.~Hohm and D.~Marques,
  ``Perturbative Double Field Theory on General Backgrounds,''
  Phys.\ Rev.\ D {\bf 93}, no. 2, 025032 (2016)
  doi:10.1103/PhysRevD.93.025032
  [arXiv:1512.02658 [hep-th]].
  
\bibitem{Hohm:2010pp} 
  O.~Hohm, C.~Hull and B.~Zwiebach,
  ``Generalized metric formulation of double field theory,''
  JHEP {\bf 1008}, 008 (2010)
  doi:10.1007/JHEP08(2010)008
  [arXiv:1006.4823 [hep-th]].
  
\bibitem{Andriot:2011uh} 
  D.~Andriot, M.~Larfors, D.~Lust and P.~Patalong,
  ``A ten-dimensional action for non-geometric fluxes,''
  JHEP {\bf 1109}, 134 (2011)
  doi:10.1007/JHEP09(2011)134
  [arXiv:1106.4015 [hep-th]].
  D.~Andriot, O.~Hohm, M.~Larfors, D.~Lust and P.~Patalong,
  ``A geometric action for non-geometric fluxes,''
  Phys.\ Rev.\ Lett.\  {\bf 108}, 261602 (2012)
  doi:10.1103/PhysRevLett.108.261602
  [arXiv:1202.3060 [hep-th]].
  D.~Andriot, O.~Hohm, M.~Larfors, D.~Lust and P.~Patalong,
  ``Non-Geometric Fluxes in Supergravity and Double Field Theory,''
  Fortsch.\ Phys.\  {\bf 60}, 1150 (2012)
  doi:10.1002/prop.201200085
  [arXiv:1204.1979 [hep-th]].

\bibitem{Hohm:2011nu} 
  O.~Hohm and S.~K.~Kwak,
  ``N=1 Supersymmetric Double Field Theory,''
  JHEP {\bf 1203}, 080 (2012)
  doi:10.1007/JHEP03(2012)080
  [arXiv:1111.7293 [hep-th]].
  I.~Jeon, K.~Lee and J.~H.~Park,
  ``Supersymmetric Double Field Theory: Stringy Reformulation of Supergravity,''
  Phys.\ Rev.\ D {\bf 85}, 081501 (2012)
  Erratum: [Phys.\ Rev.\ D {\bf 86}, 089903 (2012)]
  doi:10.1103/PhysRevD.86.089903, 10.1103/PhysRevD.85.081501, 10.1103/PhysRevD.85.089908
  [arXiv:1112.0069 [hep-th]].

\bibitem{Hohm:2010xe} 
  O.~Hohm and S.~K.~Kwak,
  ``Frame-like Geometry of Double Field Theory,''
  J.\ Phys.\ A {\bf 44}, 085404 (2011)
  doi:10.1088/1751-8113/44/8/085404
  [arXiv:1011.4101 [hep-th]].
  
\bibitem{Copland:2011wx} 
  N.~B.~Copland,
  ``A Double Sigma Model for Double Field Theory,''
  JHEP {\bf 1204}, 044 (2012)
  doi:10.1007/JHEP04(2012)044
  [arXiv:1111.1828 [hep-th]].
  D.~S.~Berman and D.~C.~Thompson,
  ``Duality Symmetric Strings, Dilatons and O(d,d) Effective Actions,''
  Phys.\ Lett.\ B {\bf 662}, 279 (2008)
  doi:10.1016/j.physletb.2008.03.012
  [arXiv:0712.1121 [hep-th]].
  D.~S.~Berman, N.~B.~Copland and D.~C.~Thompson,
  ``Background Field Equations for the Duality Symmetric String,''
  Nucl.\ Phys.\ B {\bf 791}, 175 (2008)
  doi:10.1016/j.nuclphysb.2007.09.021
  [arXiv:0708.2267 [hep-th]].
  
\bibitem{Park:2016sbw} 
  J.~H.~Park,
  ``Green-Schwarz superstring on doubled-yet-gauged spacetime,''
  JHEP {\bf 1611}, 005 (2016)
  doi:10.1007/JHEP11(2016)005
  [arXiv:1609.04265 [hep-th]].
  C.~D.~A.~Blair, E.~Malek and A.~J.~Routh,
  ``An $O(D, D)$ invariant Hamiltonian action for the superstring,''
  Class.\ Quant.\ Grav.\  {\bf 31}, no. 20, 205011 (2014)
  doi:10.1088/0264-9381/31/20/205011
  [arXiv:1308.4829 [hep-th]].

\bibitem{Jurco:2013upa} 
  B.~Jurco, P.~Schupp and J.~Vysoky,
  ``On the Generalized Geometry Origin of Noncommutative Gauge Theory,''
  JHEP {\bf 1307}, 126 (2013)
  doi:10.1007/JHEP07(2013)126
  [arXiv:1303.6096 [hep-th]].

\bibitem{Ma:2015yma} 
  C.~T.~Ma,
  ``Boundary Conditions and the Generalized Metric Formulation of the Double Sigma Model,''
  Nucl.\ Phys.\ B {\bf 898}, 30 (2015)
  doi:10.1016/j.nuclphysb.2015.06.019
  [arXiv:1502.02378 [hep-th]].

\bibitem{Ma:2014vqm} 
  C.~T.~Ma,
  ``One-Loop $\beta$ Function of the Double Sigma Model with Constant Background,''
  JHEP {\bf 1504}, 026 (2015)
  doi:10.1007/JHEP04(2015)026
  [arXiv:1412.1919 [hep-th]].

\bibitem{Lee:2015qza} 
  K.~Lee,
  ``Towards Weakly Constrained Double Field Theory,''
  Nucl.\ Phys.\ B {\bf 909}, 429 (2016)
  doi:10.1016/j.nuclphysb.2016.05.015
  [arXiv:1509.06973 [hep-th]].

\bibitem{Ma:2016vgq} 
  C.~T.~Ma and F.~Pezzella,
  ``Supergravity with Doubled Spacetime Structure,''
  Phys.\ Rev.\ D {\bf 95}, no. 6, 066016 (2017)
  doi:10.1103/PhysRevD.95.066016
  [arXiv:1611.03690 [hep-th]].

\bibitem{Kamani:2001yd} 
  D.~Kamani,
  ``T duality and noncommutative DBI action,''
  Mod.\ Phys.\ Lett.\ A {\bf 17}, 237 (2002)
  doi:10.1142/S021773230200645X
  [hep-th/0107184].

\end{thebibliography}
\end{document}